%% file: main.tex
\documentclass[conference, 10pt]{IEEEtran}

\usepackage{import}
\subimport{Praemble/}{hft-paper-praemble-main.tex}
\usepackage{comment}
\begin{document}
%
\title{On the Spectral Behavior and Normalization of a Resonance-Free and High-Frequency Stable Integral Equation}


\author{
    \IEEEauthorblockN{Tiffany L. Chhim\IEEEauthorrefmark{2}*, Simon B. Adrian\IEEEauthorrefmark{3}, and Francesco P. Andriulli\IEEEauthorrefmark{2}}
    \IEEEauthorblockA{
    \IEEEauthorrefmark{2}Department of Electronics and Telecommunications, Politecnico di Torino, Turin, Italy \\ 
    \IEEEauthorrefmark{3}Department of Electrical and Computer Engineering, Technical     University of Munich, Munich, Germany}
}

\maketitle
\begin{abstract}
The \ac{CFIE} used for solving scattering and radiation problems, although a resonance-free formulation, suffers from an ill-conditioning that strongly depends on the frequency and discretization density, both in the low- \emph{and} high-frequency regime, resulting in slow convergence rates for iterative solvers. 
This work presents a new preconditioning scheme for the \ac{CFIE} that cures the low- and the high-frequency as well as the dense discretization breakdown.
The new preconditioner for the \ac{CFIE} is based on a spherical harmonics analysis and the proper regularization with Helmholtz-type operators.
Numerical results have been obtained to prove the effectiveness of this new formulation in real scenarios.
\end{abstract}
\let\thefootnote\relax\footnote{Email addresses: tiffany.chhim@polito.it (T. L. Chhim), simon.adrian@tum.de (S. B. Adrian), francesco.andriulli@polito.it (F. P. Andriulli)}

\acresetall
\IEEEpeerreviewmaketitle

\section{Introduction}
The \ac{CFIE}, which is commonly used to solve scattering and radiation problems, suffers from the low-frequency, the dense-discretization, and the high-frequency breakdowns: the condition number of the system matrix, when the frequency is decreased (low-frequency breakdown, mesh is unchanged), when the mesh is refined (dense-discretization breakdown, frequency is unchanged), or when the frequency is increased (high-frequency breakdown) while keeping the average edge length $h$ of the mesh a fixed fraction of the wavelength (typically $h \approx \lambda/10$).

In fact, the \ac{CFIE} inherits these defects from the \ac{EFIE} and \ac{MFIE} operators, which are combined to the \ac{CFIE}, a combination which is necessary since each of these two operators alone suffers from interior resonances leading to nonphysical solutions. The low-frequency breakdown of the \ac{EFIE} has been solved first many decades ago \cite{wilton_improving_1981}, and the dense-discretization breakdown was solved first with the advent of Calderón preconditioning \cite{buffa_dual_2007}. While the high-frequency breakdown was typically tackled with algebraic preconditioners, only recently schemes have been presented which strive to obtain a bounded condition number for the asymptotic limit $k\rightarrow \infty$, with $k h$ constant, where $k$ is the wave number.

The high-frequency breakdown has been typically tackled with algebraic preconditioners, for example, based on incomplete LU decompositions or sparse approximate inverses. While these methods can reduce the condition number and thus the number of iterations required by an iterative solver to converge, they cannot prevent that the condition number grows unboundedly for the asymptotic limit. After the first seminal contributions in \cite{darbas2004preconditionneurs}, more recent contributions have been presented which obtain obtain a bounded condition number in the asymptotic limit for specific geometries (see \cite{boubendir_well-conditioned_2014,andriulli2015high} and references therein). Previous approaches have been leveraging on either square roots of differential operators or on variations of Calderon identities results. This however, resulted in either the need to use operator square root approximations or in the need for new dense operator discretizations. At the same time, the importance, for the harmonic subspace treatment, of symmetrizing both the electric and the magnetic part has been shown \cite{andriulli2015high}.  

 This work presents a novel strategy to the high-frequency preconditioning of already resonance-free equations. We leverage on the full Helmholtz operator properly combined with layer potential and with a symmetrizing factor on the magnetic part. The result is an effective approach resulting in minimal overhead with respect to a non-preconditioned scenario and which will be compatible with an harmonic regularization. Our equation will be resonant-free and will exhibit a low condition number that stays constant at all frequencies and any mesh discretization. In this paper we apply the strategy to a 2D case where we show its effectiveness via an exact eigenvalue analysis.

\section{Background and Notation}
In the following we consider equations for TM polarization; results for TE polarization can be obtained analogously. Consider a closed surface $\Gamma$, the \ac{EFIE} operator
\begin{equation}
    \vecop T_k (\veg{J}) \coloneq \frac{k \eta}{4} \int_\Gamma \veg{J}(\veg{r}') \, H_0^{(2)}(k \abs{\bm{r}-\bm{r}'}) \, \dd S(\veg r')
\end{equation}
and the \ac{MFIE} operator
\begin{multline}
    (\vecop I/2 + \vecop K_{k}) (\bm{J}) \coloneq  \veg J(\veg r)/2  \\
   - \jm/4\, \n (\veg r) \times \int_\Gamma \veg{J}(\veg{r}') \, \nabla H_0^{(2)}(k\abs{\veg{r}-\veg{r}'}) \, \dd S(\veg r')\, ,
\end{multline}
where $\eta = \sqrt{\mu/\epsilon}$ is the wave impedance, $H_0^{(2)}$ is the Hankel function of second kind, $\vecop I$ is the identity operator and $\n(\veg r)$ is the outward unit normal to $\Gamma$ at point $\veg{r}$.

Combining the \ac{EFIE} and \ac{MFIE} operator yields the \ac{CFIE} operator
\begin{equation}
    \alpha \vecop{T}_{k} + (1 - \alpha) \, \eta \, (\vecop I/2 + \vecop K_k)\, ,
\end{equation}
where the combination parameter $\alpha \in (0,1)$ (typically chosen to be \num{0.5}). The choice made in this conference paper of dealing with 2D TM equations allows us to deliver a \ac{CFIE} which in addition of being resonance-free is also low-frequency and refinement stable. In this way we can focus on the remaining high-frequency ill-conditioning.

\section{The New Formulation}

\subsection{Treatment of the EFIE and of the MFIE}

The integrative property of the TM-EFIE, which causes bad conditioning, is usually cancelled by multiplying with a derivative factor such as the TE-EFIE, which corresponds to the Calder\'on identities. Here instead, we will first multiply the integrative operator by a modified Helmholtz operator
\begin{equation}
   \vecop H_{k_\text{mod}} (\veg{J}) = \left(k_\text{mod}^{2} + \Deltaup_\Gamma \right) (\bm{J})\, ,
\end{equation}
where $\Deltaup_\Gamma$ is the surface Laplacian and where we chose $k_\text{mod} = - k - \jm 0.4 k^{1/3} R^{- 2/3}$ following \cite{darbas2004preconditionneurs} with $R$ the radius of the smallest circle encompassing the geometry $\Gamma$. This step will change the nature of the overall product into a derivative one.  We then finalize the preconditioning by a final  multiplication by the initial TM-EFIE operator obtaining
\begin{equation}
    \vecop{T}_{k_\text{mod}} \, \vecop H_{k_\text{mod}} \, \vecop{T}_{k}\, .
\end{equation}
Dually, the \ac{MFIE} is left-multiplied as follows
\begin{equation}
    (\vecop I/2 - \vecop K_{k_\text{mod}}) \, (\vecop I/2 + \vecop K_k)\, .
\end{equation}
\subsection{Novel CFIE formulation}
The complete equation we propose is obtained by combining the left-multiplied EFIE and the MFIE  to get
\begin{equation}
    c_\text{M} \, \eta^{2} \, (\vecop I/2 - \vecop{K}_{k_\text{mod}}) \, (\vecop I/2 + \vecop K_k) + c_\text{E} \, \vecop{T}_{k_\text{mod}} \, \vecop H_{k_\text{mod}} \, \vecop{T}_{k}\, ,
\end{equation}
where $c_\text{M}$ and $c_\text{E}$ are normalization factors. A spectral asymptotic study has been made in order to normalize the equation and achieve an asymptotic value of \num{1} for the spectrum.  In order to balance both terms, we seek to obtain a limit of 0.5 for each term in the elliptic regime. This leads to the following normalizing factors for the \ac{MFIE} and \ac{EFIE} terms respectively 
\begin{equation}
   c_\text{M}=\frac{4}{\eta^{2}},\quad   c_\text{E}=\frac{4}{k \, k_\text{mod} \, \eta^{2} R^{2}}\, .
\end{equation}

\section{Numerical Results}

As a benchmark case we have used a 2D circle of radius $R = 1$. Fig.~\ref{fig:standard} corresponds to the spectrum of the standard \ac{CFIE} at varying frequencies and clearly shows that, although the equation is resonant-free, the  overall conditioning still increases as a function of the frequency. The spectrum of our new equation is displayed in Fig. \ref{fig:magic}. On top of being resonance-free and low-frequency and refinement stable, it is also stable in the high-frequency regime.
\begin{figure}
\centering
    \input{./Figures/SpectrumStandardCFIE}
    \caption{Circle: spectrum of the unpreconditioned \ac{CFIE}.}
    \label{fig:standard}
\end{figure}
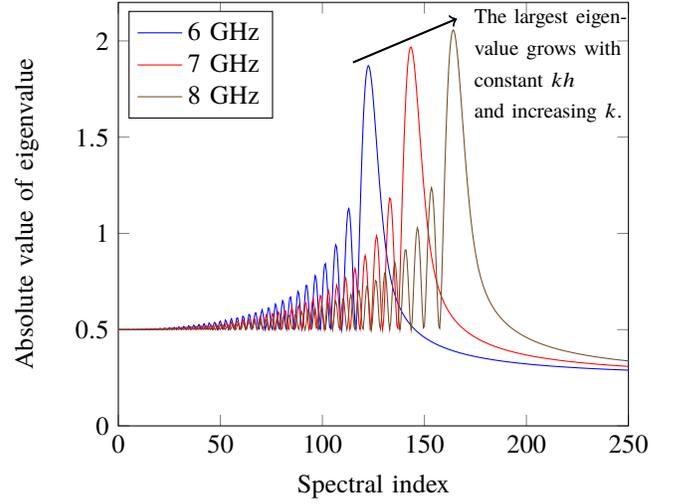
\begin{figure}
\centering
  \input{./Figures/SpectrumMagicCFIE}
  \caption{Circle: spectrum of the new \ac{CFIE} formulation.}
  \label{fig:magic}
\end{figure}
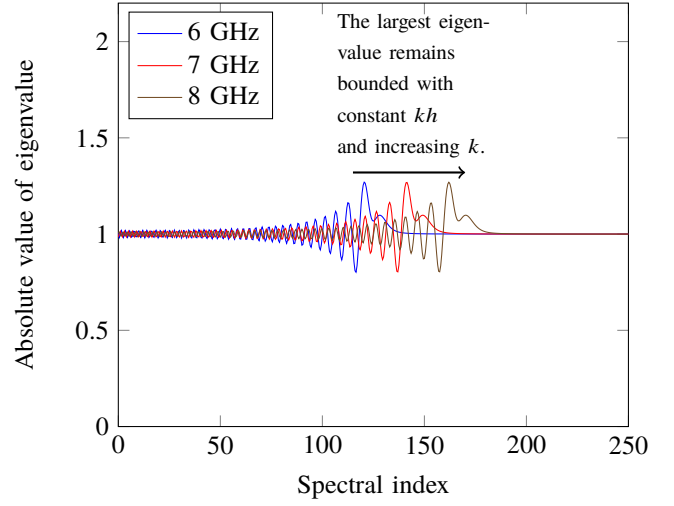

This work was supported by the European Research Council (ERC) under the European Union’s Horizon 2020 research and innovation programme (grant agreement No 724846, project 321).

\bibliographystyle{IEEEtran}
\bibliography{references}

\end{document}

%% file: Praemble/hft-paper-praemble-main.tex
\input{hft-paper-praemble-packages.tex}
\input{hft-macros.tex}

\input{sba-macros.tex}

%% file: Praemble/hft-paper-praemble-packages.tex

\usepackage[utf8]{inputenc}

\usepackage[T1]{fontenc}

\usepackage{graphicx}
\usepackage{xcolor}

\usepackage{newtxtext}
\usepackage{amsthm}
\usepackage[slantedGreek]{newtxmath}
\usepackage[OMLmathsfit]{isomath}
\DeclareMathAlphabet{\mathbfsf}{\encodingdefault}{\sfdefault}{bx}{n}
\usepackage{bm}

\usepackage{envmath}
\usepackage{mathtools}  
\usepackage{siunitx}
\usepackage{amssymb}

\usepackage[caption=false,font=footnotesize]{subfig}

\usepackage{booktabs}
\usepackage{footmisc}  

\usepackage{url}

\input{hft-packages-theorem.tex}

\input{hft-packages-review.tex}
\input{hft-packages-tikz-pgfplots.tex}
\input{hft-packages-acro.tex}
\input{hft-packages-cleveref.tex}

%% file: Praemble/hft-packages-theorem.tex
\theoremstyle{definition}

\theoremstyle{plain}

\theoremstyle{remark}

%% file: Praemble/hft-packages-review.tex
\usepackage{lineno}
\modulolinenumbers[5]
\usepackage{todonotes}
\usepackage{umoline}


%% file: Praemble/hft-packages-tikz-pgfplots.tex
\usepackage{pgfplots}
\usepackage{pgfplotstable}
\newlength\figureheight
\newlength\figurewidth

%% file: Praemble/hft-packages-acro.tex
\usepackage{acro}

\DeclareAcronym{ACA}
{
    short = ACA ,
    long = adaptive cross approximation
}

\DeclareAcronym{MLFMM}
{
    short = MLFMM ,
    long = multilevel fast mutipole method
}

\DeclareAcronym{RCS}
{
    short = RCS ,
    long = bistatic radar cross section
}

\DeclareAcronym{EFIE}
{
    short =  EFIE ,
    long = electric field integral equation
}

\DeclareAcronym{SPD}
{
    short =  SPD ,
    long = {symmetric, positive definite}
}

\DeclareAcronym{SPSD}
{
    short =  SPD ,
    long = {symmetric, positive semi-definite}
}

\DeclareAcronym{PEC}
{
    short =  PEC ,
    long = perfectly electrically conducting
}

\DeclareAcronym{RWG}
{
    short = RWG ,
    long = Rao-Wilton-Glisson
} 

\DeclareAcronym{BC}
{
    short = BC ,
    long = Buffa-Christiansen
}

\DeclareAcronym{SVD}
{
    short = SVD ,
    long = singular value decomposition
} 

\DeclareAcronym{GMRES}
{
    short = GMRES ,
    long = generalized minimal residual method
}

\DeclareAcronym{CGS}
{
    short = CGS ,
    long = conjugate gradient squared
} 

\DeclareAcronym{RFCMP}
{
    short = RFCMP ,
    long = refinement-free Calderón multiplicative preconditioner
}

\DeclareAcronym{WC}
{
    short = WC ,
    long = Wilton-Chen ,
} 

\DeclareAcronym{MoM}
{
    short = MoM ,
    long = method of moments ,
}

\DeclareAcronym{CFIE}
{
    short = CFIE ,
    long = combined field integral equation ,
} 

\DeclareAcronym{MFIE}
{
    short = MFIE ,
    long = magnetic field integral equation ,
} 

%% file: Praemble/hft-packages-cleveref.tex
\usepackage[english]{cleveref}

\Crefname{defn}{definition}{definitions}
\Crefname{defn}{Definition}{Definitions}

\Crefname{asm}{assumption}{assumptions}
\Crefname{asm}{Assumption}{Assumptions}

\crefname{lem}{lemma}{lemmas} 
\Crefname{lem}{Lemma}{Lemmas}

\crefname{prop}{proposition}{propositions} 
\Crefname{prop}{Proposition}{Propositions}

\crefname{thm}{theorem}{theorms} 
\Crefname{thm}{Theorem}{Theorms}

\crefname{cor}{corollary}{corollaries}
\Crefname{cor}{Corollary}{Corollaries}

%% file: Praemble/hft-macros.tex
\input{hft-macros-environments.tex} 
\input{hft-macros-maths.tex}

\input{hft-macros-booktabs.tex}


%% file: Praemble/hft-macros-environments.tex
\newcounter{subequation}
\newlength\mtabskip\mtabskip=-1.25cm

\def\mtabLong{long}
\makeatletter

\makeatother

%% file: Praemble/hft-macros-maths.tex
\newcommand{\mr}{\mathrm}

\newcommand{\veg}[1]{\bm{#1}}     
\newcommand{\vecop}[1]{\bm{\mathcal{#1}}} 


\newcommand{\n}{\hat{\bm{n}}}


\newcommand{\dd}{\mathrm{d}}  


\newcommand{\jm}{\mathrm{j}}  



\DeclarePairedDelimiter{\abs}{\lvert}{\rvert}




\newcommand\restr[2]{{
        \left.\kern-\nulldelimiterspace 
        #1 
        \vphantom{|} 
        \right|_{#2} 
}}

\newcommand\rst[3]{{
        \left.\kern-\nulldelimiterspace 
        #1 
        \vphantom{|} 
        \right|_{#2}^{#3} 
}}


%% file: Praemble/hft-macros-booktabs.tex

\newcolumntype {n}{c}
\newcolumntype {N}{>{\small}c}
\newcolumntype {L}{>{\small}l}
\newcolumntype {F}{>{\footnotesize}c}
\newcolumntype {v}[1]{>{\raggedright \hspace {0pt}} p {#1}}
\newcolumntype {V}[1]{>{\small \raggedright \hspace {0pt}} p {#1}}
\newcolumntype{d}[1]{>{\DC@{.}{.}{#1}}c<{\DC@end}}

%
\newcolumntype{R}[1]{%
    >{\begin{turn}{90}\begin{minipage}{#1}\small\raggedright\hspace{0pt}}l%
            <{\end{minipage}\end{turn}}%
}

%% file: Praemble/sba-macros.tex
\newcommand{\TA}{\vecop T_{\kern-4pt\mr{A}}}
\newcommand{\TPhi}{\vecop T_{\kern-4pt\Phiup}}

%% file: Figures/SpectrumStandardCFIE.tex
\resizebox {0.99\columnwidth} {!} {
\begin{tikzpicture}
\begin{axis}[%
xlabel={Spectral index},
ylabel={Absolute value of eigenvalue},
xmin=0,
xmax=250,
ymin=0,
ymax=2.2,
legend style={ anchor=north west,
    at={(0.02,.98)}
},
]

\pgfplotstableread[col sep=semicolon,  header=true]{./Figures/SpectrumStandardCFIE.csv}\datatable



\addplot+ [mark=none]
table[x=sindex, y=6x10_9]   from \datatable {};
\addlegendentry{6 GHz};

\addplot+ [mark=none]
table[x=sindex, y=7x10_9]   from \datatable {};
\addlegendentry{7 GHz};

\addplot+ [mark=none]
table[x=sindex, y=8x10_9]   from \datatable {};
\addlegendentry{8 GHz};

       \draw[thick,->](axis cs:115,1.888048193) -- (axis cs:166,2.115520482);
    \node[anchor=west, align=left] (source) at (axis cs:170,1.8715){\footnotesize The largest eigen-\\ \footnotesize value grows with\\ \footnotesize constant $kh$\\ \footnotesize and increasing $k$.};
       
\end{axis}
\end{tikzpicture}%
}

%% file: Figures/SpectrumMagicCFIE.tex
\resizebox {0.99\columnwidth} {!} {
\begin{tikzpicture}
\begin{axis}[%
xlabel={Spectral index},
ylabel={Absolute value of eigenvalue},
xmin=0,
xmax=250,
ymin=0,
ymax=2.2,
legend style={ anchor=north west,
    at={(0.02,.98)}
},
]

\pgfplotstableread[col sep=semicolon,  header=true]{./Figures/SpectrumMagicCFIE.csv}\datatable



\addplot+ [mark=none]
table[x=sindex, y=6x10_9]   from \datatable {};
\addlegendentry{6 GHz};

\addplot+ [mark=none]
table[x=sindex, y=7x10_9]   from \datatable {};
\addlegendentry{7 GHz};

\addplot+ [mark=none]
table[x=sindex, y=8x10_9]   from \datatable {};
\addlegendentry{8 GHz};

\draw[thick,->](axis cs:115,1.32) -- (axis cs:170,1.32);
\node[anchor=south, align=left] (source) at (axis cs:145,1.35){\footnotesize The largest eigen-\\ \footnotesize value remains\\ \footnotesize bounded with\\ \footnotesize constant $kh$\\ \footnotesize and increasing $k$.};     

\end{axis}
\end{tikzpicture}%
}